# Tailored subcycle nonlinearities of ultrastrong light-matter coupling

J. Mornhinweg[1], M. Halbhuber[1], C. Ciuti[2], D. Bougeard[1], R. Huber[1,†], and C. Lange[1,3†]

[1]Department of Physics, University of Regensburg, 93040 Regensburg, Germany

[2]Université de Paris, Laboratoire Matériaux et Phénomènes Quantiques,

CNRS, F-75013 Paris, France

[3]Fakultät Physik, Technische Universität Dortmund, 44227 Dortmund, Germany

**We explore the nonlinear response of tailor-cut light-matter hybrid states in a novel regime, where both the Rabi frequency induced by a coherent driving field and the vacuum Rabi frequency set by a cavity field are comparable to the carrier frequency of light. In this previously unexplored strong-field limit of ultrastrong coupling, subcycle pump-probe and multi-wave mixing nonlinearities between different polariton states violate the normal-mode approximation while ultrastrong coupling remains intact, as confirmed by our mean-field model. We expect such custom-cut nonlinearities of hybridized elementary excitations to facilitate non-classical light sources, quantum phase transitions, or cavity chemistry with virtual photons.**

Exposing solid-state electrons to extremely strong optical fields can drive highly non-perturbative dynamics faster than an oscillation cycle of the carrier wave, generating high-harmonic or high-order sideband radiation [1–4]. For a field resonant to an electronic excitation, the Rabi frequency $\Omega_R$ quantifies the rate of periodic absorption and stimulated emission. When $\Omega_R$ exceeds the carrier frequency of light, $\omega_c$, carrier-wave Rabi flopping occurs, accompanied by strong nonlinearities [5,6].

In a complementary setting, strongly enhanced vacuum field fluctuations of sub-wavelength microcavities can be coupled to electronic excitations, causing a periodic energy exchange between the electronic transition and the cavity field at a rate called the vacuum Rabi frequency, $\Omega_R^v$. In consequence, light-matter hybrid states – cavity polaritons – emerge, with custom optical and electronic properties. When $\Omega_R^v$ approaches or even exceeds $\omega_c$, ultrastrong [7–16] and deep-strong coupling [12,17] emerge, respectively. Under such exotic conditions, anti-resonant interactions can simultaneously create (or annihilate) both a cavity photon and an electronic excitation. As a result, the ground state contains a sizable population of virtual photons. Even under equilibrium conditions examined by weak probe pulses, this limit of light-matter interaction has brought forth complex quantum effects including the vacuum Bloch-Siegert shift [13], modified electronic transport [14,18], cavity chemistry [19,20], or vacuum-field-induced superconductivity [21,22].

Combing both scenarios, intense laser fields are expected to drive strong-field nonlinearities of light-matter coupled hybrid states in a situation where $\Omega_R^v$ and $\Omega_R$ simultaneously become comparable to $\omega_c$, leading to subcycle polarization dynamics driven by both, the vacuum field, and a coherent field. Such forced interactions of normal modes would produce previously inaccessible optical and electronic states beyond the eigenstates of the light-matter coupled system, unfolding an unprecedented parameter space for exploring dynamical quantum phenomena. Moreover, nonlinearities may be exploited to tailor correlations or phase transitions, and facilitate new devices including sources of single [23] and entangled photons, or non-classical quantum states of light [24].

Here, we explore this novel strong-field limit, by driving ultrastrongly cavity-coupled Landau electrons ($\Omega_R^v/\omega_c = 0.77$) by strong, single-cycle THz pulses reaching $\Omega_R = 2\pi \times 2.7$ THz (Fig. 1a). With $\Omega_R^v \approx \omega_c \approx \Omega_R$, non-perturbative excitations of the hybrid system including coherent Landau ladder climbing by up to 10 rungs occur on subcycle time scales. Strong pump-probe and coherent four- [25], six-, and eight-wave-mixing nonlinearities occur for each polariton branch individually. Importantly, controlled nonlinear interactions between individual polaritons arise, dynamically breaking the normal-mode approximation. Our theory quantitatively reproduces these data and links the nonlinearities to dynamical Coulomb correlations during non-perturbative excitation of Landau electrons. This first exploration of strong-field interactions of normal modes of ultrastrong coupling paves the way for designing light-matter hybrid states with custom-cut nonlinear functionalities.

Our structures couple the eigenmodes of metallic THz resonators to the cyclotron resonance (CR) of two-dimensional electron gases hosted in 6 GaAs quantum wells (QWs) of a thickness of 10 nm, separated by AlGaAs barriers (Fig. 1a). Each QW is n-doped with an electron density of $1.75 \times 10^{12} \text{cm}^{-2}$. A variable magnetic bias field $B$ oriented perpendicularly to the surface Landau-quantizes the electrons and sets the CR frequency, $\nu_c$. The resonators are fabricated on top of the heterostructure by electron-beam lithography. Single-cycle THz waveforms (Fig. 1a, red waveform) probing the transmission spectrum of the bare resonators reveal a fundamental LC mode at a frequency of $\nu_{LC} = \omega_{LC}/2\pi = 0.81$ THz, and a dipolar mode at $\nu_{DP} = 1.8$ THz. Including the QWs, the electric field of these modes ultrastrongly couples to the polarization of the CR. As $\nu_c$ is varied, the transmission spectrum reveals several minima corresponding to three pronounced light-matter coupled modes (Fig. 1b). A lower polariton (LP$_1$) branch emerges from the CR at low frequencies, while a corresponding upper polariton resonance (UP$_1$) is observed at a frequency of 1.5 THz for $\nu_c = 0$ THz. As $\nu_c$ is increased, the polariton frequencies grow with opposite curvatures, causing the characteristic anti-crossing shape. Off-resonant coupling of the CR to the dipolar cavity mode creates a third resonance labelled UP$_2$, located $\approx 0.2$ THz below the UP$_1$ branch (see Supplemental Material). THz absorption by the CR in uncoupled areas between resonator structures explains the transmission minimum at $\nu_c$ (dark blue region). Diagonalizing the light-matter Hamiltonian for the LP$_1$ and UP$_1$ resonances, we obtain a coupling strength of $\Omega_R^v/\omega_{LC} = 0.77$ (see Supplemental Material).

We interrogate the system's nonlinear response by field-resolved two-dimensional spectroscopy [26–29]. This comprehensive characterization strategy identifies all multi-photon interaction pathways by their characteristic Liouville path, even in the presence of much stronger linear or nonlinear contributions. Two single-cycle THz waveforms labelled A and B with a relative time delay $\tau$ and peak amplitudes of 1.3 kV/cm and 2.5 kV/cm, respectively, generated by tilted-pulse front optical rectification [30], are collinearly focused through the structure (Fig. 2a) and subsequently detected by electro-optic sampling.

Mechanical choppers enable us to measure the transmitted fields $\mathcal{E}_A$ and $\mathcal{E}_B$ of both pulses individually, as well as the combined field $\mathcal{E}_{AB}$, from which we determine the nonlinear response, $\mathcal{E}_{nl} = \mathcal{E}_{AB} - \mathcal{E}_A - \mathcal{E}_B$.

First, we switch off the magnetic field ($v_c = 0$ THz) and investigate the nonlinearities of the cavity coupled to the electron plasma, where the relevant UP$_1$ and UP$_2$ modes assume frequencies of 1.5 and 1.2 THz, respectively (Fig. 1b). The nonlinear response $\mathcal{E}_{nl}$ displays a rich time-domain structure with multiple oscillation patterns (Fig. 2b). Fronts of constant phase of pulse A appear as vertical lines, while those of pulse B manifest as diagonal lines for which $t - \tau$ is constant. $\mathcal{E}_{nl}$ is strongest around $\tau = 0$ ps, where both pulses interfere constructively, causing an oscillation with the frequencies of both upper polaritons. The resulting pronounced beating along the $t$ axis periodically minimizes the envelope of $\mathcal{E}_{nl}$ at $t \cong 3$ ps and 5.5 ps. For $\tau = \pm 0.25$ ps, destructive interference of the pulses minimizes the THz peak amplitude and, therefore, $\mathcal{E}_{nl}$.

Next, we set $B$ to 2.3 T and measure $\mathcal{E}_{nl}$ at the anti-crossing point, $v_c \approx v_{LC}$, where the LP$_1$, UP$_1$ and UP$_2$ resonances are centered at frequencies of 0.4 THz, 1.65 THz, and 1.3 THz, respectively (Fig. 1b). Now, a higher level of coherence is apparent by the slower decay of $\mathcal{E}_{nl}$, leading to a finite amplitude for delay times up to $\tau = 5$ ps (Fig. 2c). Moreover, here, the beating pattern is caused by interactions of three rather than two coupled resonances. The resulting complex nature of $\mathcal{E}_{nl}$ gives reason to expect a rich structure of nonlinearities.

For a systematic decomposition of all nonlinear optical interaction processes, we perform a two-dimensional Fourier transform [26–29]. The resulting nonlinear amplitude spectrum, $\mathcal{A}_{nl}(v_t, v_\tau)$, depends on the frequencies $v_t$ and $v_\tau$ associated with the respective delay times. For $v_\tau = 0$ THz, where $\mathcal{E}_{nl}$ is averaged along the $\tau$ axis, $\mathcal{A}_{nl}$ reveals three distinct, narrowband signatures at the polariton frequencies (Fig. 3a). Generally, each maximum of $\mathcal{A}_{nl}$ corresponds to a distinct multi-photon process combining the pseudo-wave vectors of the incident fields, $\mathbf{k}_A = (v_j, 0)$ and $\mathbf{k}_B = (v_j, v_j)$, where $v_j$ with $j \in \{LP_1, UP_1, UP_2\}$ denotes the polariton frequencies, and off-resonant spectral components can be neglected [26–29]. Peaks along the black dashed lines mark pump-probe (PP) processes resulting from third-order nonlinearities, $P_{\text{PP}}^{(3)}(\nu_j) = \chi^{(3)}(\nu_j, \nu_j, -\nu_j)\mathcal{A}_m(\nu_j)\mathcal{A}_n(\nu_j)\mathcal{A}_n(-\nu_j)$, whereby $\chi^{(3)}$ is the corresponding nonlinear susceptibility, and $\mathcal{A}_m$ with $m, n \in \{A, B\}, m \neq n$, denotes the spectral amplitudes of the incident fields. The resulting wave vectors, $\mathbf{k}_{\text{PP1}} = \mathbf{k}_A + \mathbf{k}_B - \mathbf{k}_B$ and $\mathbf{k}_{\text{PP2}} = \mathbf{k}_B + \mathbf{k}_A - \mathbf{k}_A$, each cancel one of the incident field components, showing that the PP processes probe the population response. In contrast, four-wave mixing (4WM) signals along the red dashed lines depend on the phases of both fields and probe the nonlinear polarization, $P_{\text{4WM}}^{(3)}(\nu_j) = \chi^{(3)}(\nu_j, \nu_j, -\nu_j)\mathcal{A}_m(\nu_j)\mathcal{A}_m(\nu_j)\mathcal{A}_n(-\nu_j)$. We illustrate this wave vector decomposition of third-order nonlinearities (Fig. 3b) for the 4WM signals located at $(\nu_j, -\nu_j)$ in Fig. 3c, where blue, red and green arrows mark the Liouville paths for the LP$_1$, UP$_2$ and UP$_1$ resonances, respectively. In each case, $P_{\text{4WM}}^{(3)}(\nu_j)$ requires absorbing two photons from $\mathcal{E}_A$ (horizontal arrows), and emitting one photon into $\mathcal{E}_B$ (diagonal arrows), creating a polarization at a frequency of $\nu_j$ which subsequently reemits into the far-field, where it is detected electro-optically. All photons are resonant to the respective polariton. Equivalent signatures, yet with up to eight-wave mixing processes, are obtained for a reference sample containing a single QW (Supplemental Material).

Intriguingly, $\mathcal{A}_{\text{nl}}$ reveals additional maxima at frequencies such as $(\nu_{\text{LP}_1}, -\nu_{\text{UP}_2})$ resulting from nonlinear polarization mixing between different polariton resonances (Fig. 3d). Here, $\mathcal{E}_A$ contributes a photon resonant to LP$_1$ (blue arrow) and one resonant to UP$_2$ (red horizontal arrow), while a photon resonant to UP$_2$ is emitted into $\mathcal{E}_B$ (red diagonal arrow), driving a nonlinear polarization $P_{\text{4WM}}^{(3)}(\nu_{\text{LP}_1}) = \chi^{(3)}(\nu_{\text{LP}_1}, \nu_{\text{UP}_2}, -\nu_{\text{UP}_2})\mathcal{A}_A(\nu_{\text{LP}_1})\mathcal{A}_A(\nu_{\text{UP}_2})\mathcal{A}_B(-\nu_{\text{UP}_2})$. Equivalent 4WM and PP processes mixing the polarizations of UP$_1$ and UP$_2$ are highlighted in Figs. 3e,f, respectively. These inter-polariton signatures mark a strong-field limit of ultrastrong coupling in which the normal-mode approximation is violated. The broad variety of nonlinear interactions resulting from all possible wave vector combinations (see Supplemental Material) underscores the high design flexibility regarding custom nonlinearities of light-matter hybrid states by subwavelength design of ultrastrong coupling.

The observation of nonlinear interactions of independent normal modes calls for a quantitative theoretical description implementing the coupling of the resonator fields to the electronic polarization of the multi-level Landau fan. In second quantization, our Hamiltonian

$$\hat{H} = \sum_j \hbar\omega_j \hat{a}_j^\dagger \hat{a}_j + \hbar\omega_c \hat{b}^\dagger \hat{b} + \hbar \sum_j \Omega_{R,j}^v (\hat{a}_j + \hat{a}_j^\dagger)(\hat{b} + \hat{b}^\dagger) + \hbar \sum_j D_j (\hat{a}_j + \hat{a}_j^\dagger)^2 + \hat{H}_{\text{ext}} \quad (1)$$

contains bosonic annihilation operators, $\hat{a}_j$, and frequencies, $\omega_j$, for the j-th cavity mode, and their cyclotron electronic counterparts, $\hat{b}$ and $\omega_c$. The respective bare energies motivate the first two terms. The third term couples the cavity modes to the electronic polarization by vacuum Rabi frequencies $\Omega_{R,j}^v$ including anti-resonant interactions, $\hat{a}_j \hat{b} + h.c.$. The fourth term treats the diamagnetic shift of the cavity modes, $D_j = {\Omega_{R,j}^v}^2/\omega_c$. Finally, $\hat{H}_{\text{ext}}$ couples the cavity to the far field.

We describe our coherent spectroscopy by a mean-field approach and obtain the polarization dynamics by Heisenberg's equation of motion for the expectation values of the operators. A comparison with the experimental data allows for calibrating $\Omega_{R,j}^v$, damping terms, and the Rabi frequency associated with the external field, $\Omega_R = 2\pi \times 2.7$ THz (Supplemental Material). While this bosonic description is valid for weak fluences, strong-field excitation unravels the fermionic nature of the electronic system. We implement the full complexity of the multi-level Landau fan by a density matrix formalism which considers the coherent polarization and incoherent populations of $10^2$ individual Landau levels (LLs). The density matrix $\rho$ contains the LL populations $n_l = \rho_{ll}$ on the diagonal and the inter-LL polarizations in the adjacent minor diagonal, $p_l = \rho_{l,l+1} = \rho_{l+1,l}^*$. The total polarization $\beta = \sum_l p_l$ is coupled back to the cavity. The dynamics of $\rho$ is obtained by solving the von Neumann equation. Dephasing and relaxation processes driving the system back towards the equilibrium state before excitation, $\rho_0$, are implemented by a corresponding matrix $\gamma^L$ (see Supplemental Material). Figure 4a shows the evolution of the population of each LL for the situation of Fig. 2c. Within the FWHM of $\mathcal{E}_B$ of 2 ps around $t = 0$ ps, Landau electrons are coherently excited from up to 5 rungs below to 5 rungs above the Fermi level. The maximum energy spread remains slightly below the energy of the longitudinal optical phonon of GaAs. This threshold is exceeded by doubling $\mathcal{E}_B$, leading to rapid electron-phonon scattering and strong dephasing (Supplemental Material). We condense the complex system state into an excitation parameter, $\rho_{\text{exc}}$, which quantifies the number of Landau excitations relative to thermal equilibrium, and corresponds to $|\beta|^2$ of the bosonic Hamiltonian, without nonlinearities (Supplemental Material). After excitation, the periodic energy exchange of the cavity and the Landau electrons causes periodically alternating local maxima of $\rho_{\text{exc}}$ and $|\alpha_j|$ (Figs. 4b,c).

Moreover, our model explains the microscopic origin of the nonlinearities. A first potential mechanism is given by the non-parabolic conduction band of GaAs, which causes a non-equidistant progression of the LL frequencies, $\omega_l = \sqrt{\omega_{\text{np}}^2/4 + \omega_{\text{np}} eB/m^*(l + 0.5)}$ [31]. Here, $m^* = 0.066\, m_{\text{e}}$ is the effective mass in the parabolic approximation, $\omega_{\text{np}} = 2\pi \times 237$ THz is a fit parameter quantifying quartic contributions to the band curvature, and $m_{\text{e}}$ is the electron mass. Although qualitatively matching the experiment, the calculated dynamics (Fig. 4d,e) progress much slower, leading to a red-shifted spectrum dominated by the LP$_1$ resonance. Moreover, the peak amplitude of $\mathcal{A}_{\text{nl}}$ reaches only 8% of the experimentally observed value, while distinctly surpassing the noise floor, which amounts to 1.2% of the experimental peak amplitude of $\mathcal{A}_{\text{nl}}$ (Supplemental Information). This evidences the presence of additional, strong nonlinear contributions.

While gold resonators are expected to respond linearly up to much stronger fields [32], the CR has been shown to exhibit nonlinearities beyond Kohn's theorem by dynamical Coulomb correlations at field amplitudes of a few kV/cm [29]. To first order, the resulting renormalization of the LL energies and the inter-LL dipole moments depend on the instantaneous degree of excitation, $\rho_{\text{exc}}(t)$. Scaling both effects for a quantitative fit with the experiment (Supplemental Material), the full calculation (Fig. 4f) reproduces the rapidly oscillating field components originating from the UP resonances, and the slower modulation induced by the LP$_1$ resonance. Moreover, relative and absolute spectral amplitudes of pump-probe and four-wave mixing signals are adequately rendered for the individual polaritons, and for the nonlinear correlations between separate polariton resonances (Fig. 4g). A switch-off analysis shows that only the combination of all nonlinearities enables this quantitative agreement (Supplemental Material). Our theory describes the creation of distinct resonances by strong-field interactions of light-matter hybrid modes which greatly expands the parameter space for tailoring deep-strong coupling – a regime that was hitherto restricted to linear interactions with weak probe fields.

In conclusion, we have investigated a novel regime of ultrastrong light-matter coupling where subcycle polarization dynamics of a many-body electronic system are driven by both vacuum fluctuations and a coherent external field. Non-perturbative excitation stimulates pump-probe and coherent multi-wave mixing nonlinearities. Moreover, nonlinearities linking the polarization of separate polaritons break the normal-mode approximation dynamically, creating resonances inaccessible by linear spectroscopy. Whereas deep-strong coupling features custom-cut light-matter hybrid modes with exotic properties in equilibrium, the nonlinear interactions of eigenstates observed here open up previously inaccessible degrees of freedom of cavity quantum electrodynamics (c-QED). This strategy may unveil quantum phenomena in similarity to c-QED in the strong-coupling regime, where Bose-Einstein condensates or quantum fluids of light have been routinely created [33]. Following a similar philosophy, nonlinear optical control of deep-strong coupling could massively enhance ground-state instabilities and phase transitions in superconductivity [21,22], cavity chemistry [19,20], or vacuum-modified electronic transport [14,18], and enable the detection of quantum vacuum radiation [34]. The comparably low field amplitudes on the order of 1 kV/cm may facilitate innovative quantum devices exploiting parametric polariton nonlinearities or the generation of squeezed quantum states of light.

We thank Andreas Bayer for experimental support. We gratefully acknowledge support by the DFG through grants no. LA 3307/1-2, BO 3140/3-2, and HU 1598/2, as well as by the European Research Council (ERC) through Future and Emerging Technologies (FET) grant no. 737017 (MIR-BOSE).

## References

[1] A. H. Chin, O. G. Calderón, and J. Kono, Phys. Rev. Lett. **86**, 3292 (2001).

[2] B. Zaks, R. B. Liu, and M. S. Sherwin, Nature **483**, 580 (2012).

[3] O. Schubert *et al.*, Nat. Photonics **8**, 119 (2014).

[4] F. Langer *et al.*, Nature **533**, 225 (2016).

[5] O. D. Mücke, T. Tritschler, M. Wegener, U. Morgner, and F. X. Kärtner, Phys. Rev. Lett. **87**, 057401 (2001).

[6] J. Raab *et al.*, Opt. Express **27**, 2248 (2019).

[7] G. Günter *et al.*, Nature **458**, 178 (2009).

[8] A. A. Anappara, S. De Liberato, A. Tredicucci, C. Ciuti, G. Biasiol, L. Sorba, and F. Beltram, Phys. Rev. B **79**, 201303(R) (2009).

[9] Y. Todorov, A. M. Andrews, R. Colombelli, S. De Liberato, C. Ciuti, P. Klang, G. Strasser, and C. Sirtori, Phys. Rev. Lett. **105**, 196402 (2010).

[10] G. Scalari *et al.*, Science **335**, 1323 (2012).

[11] J. Keller, G. Scalari, S. Cibella, C. Maissen, F. Appugliese, E. Giovine, R. Leoni, M. Beck, and J. Faist, Nano Lett. **17**, 7410 (2017).

[12] A. Bayer, M. Pozimski, S. Schambeck, D. Schuh, R. Huber, D. Bougeard, and C. Lange, Nano Lett. **17**, 6340 (2017).

[13] X. Li, M. Bamba, Q. Zhang, S. Fallahi, G. C. Gardner, W. Gao, M. Lou, K. Yoshioka, M. J. Manfra, and J. Kono, Nat. Photonics **12**, 324 (2018).

[14] G. L. Paravicini-Bagliani *et al.*, Nat. Phys. **15**, 186 (2019).

[15] A. F. Kockum, A. Miranowicz, S. De Liberato, S. Savasta, and F. Nori, Nat. Rev. Phys. **1**, 19 (2019).

[16] P. Forn-Díaz, L. Lamata, E. Rico, J. Kono, and E. Solano, Rev. Mod. Phys. **91**, 025005 (2019).

[17] F. Yoshihara, T. Fuse, S. Ashhab, K. Kakuyanagi, S. Saito, and K. Semba, Nat. Phys. **13**, 44 (2017).

[18] E. Orgiu *et al.*, Nat. Mater. **14**, 1123 (2015).

See Supplemental Material for an in-depth discussion of the theory, more detail on the experimental setup and additional data.

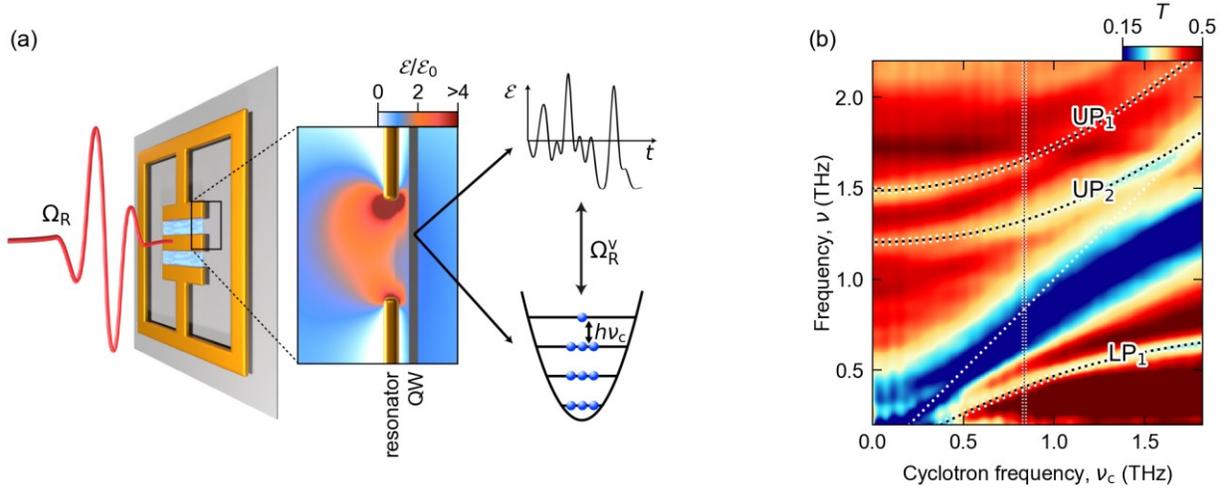

FIG. 1. Ultrastrongly light-matter coupled structure. (a) Schematic showing the THz far field (red waveform) which couples to the optical modes of the resonator (gold structure) with a strength defined by the Rabi frequency $\Omega_R$. Gray layer: quantum well structure. The magnified area shows the near field amplitude inside the resonator gap. Here, vacuum field fluctuations of the resonator mode (upper right graph) couple to the Landau levels (lower right schematic) of the quantum wells with a strength quantified by the vacuum Rabi frequency $\Omega_R^v$. (b) Measured transmission spectra as a function of the cyclotron frequency, $\nu_c$. Dotted curves: lower (LP$_1$) and upper polariton mode (UP$_1$) obtained from Hopfield's model (see Supplemental Material) for a coupling strength of $\Omega_R^v/\omega_c = 0.77$. UP$_2$: additional polariton mode coupled to the higher-order, dipolar mode of the resonator. Vertical dotted line: Anti-crossing point, $\nu_c = 0.84$ THz. Diagonal dotted line: Cyclotron resonance.

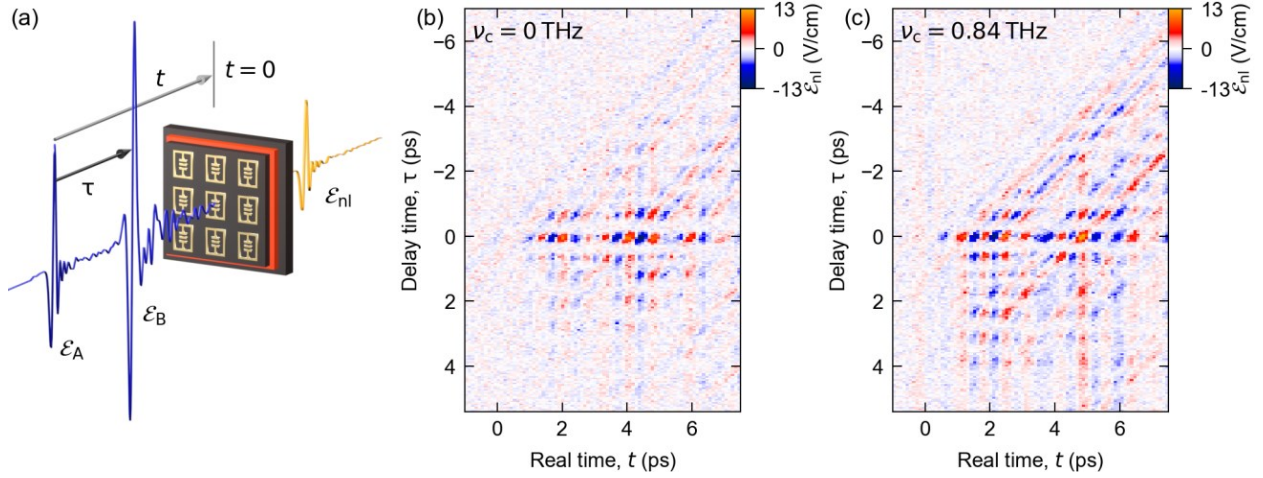

FIG. 2. (a) Experimental geometry of amplitude and field-resolved two-dimensional THz spectroscopy. The electro-optic delay time (also called real time) is denoted by $t$, whereas $\tau$ is the relative delay of the two THz waveforms. (b) Measured nonlinear response $\mathcal{E}_{nl}(t,\tau)$ as a function of $t$ and $\tau$ for $\nu_c = 0$ THz. (c) $\mathcal{E}_{nl}(t,\tau)$ at the anti-crossing point, where $\nu_c \approx \nu_{LC}$.

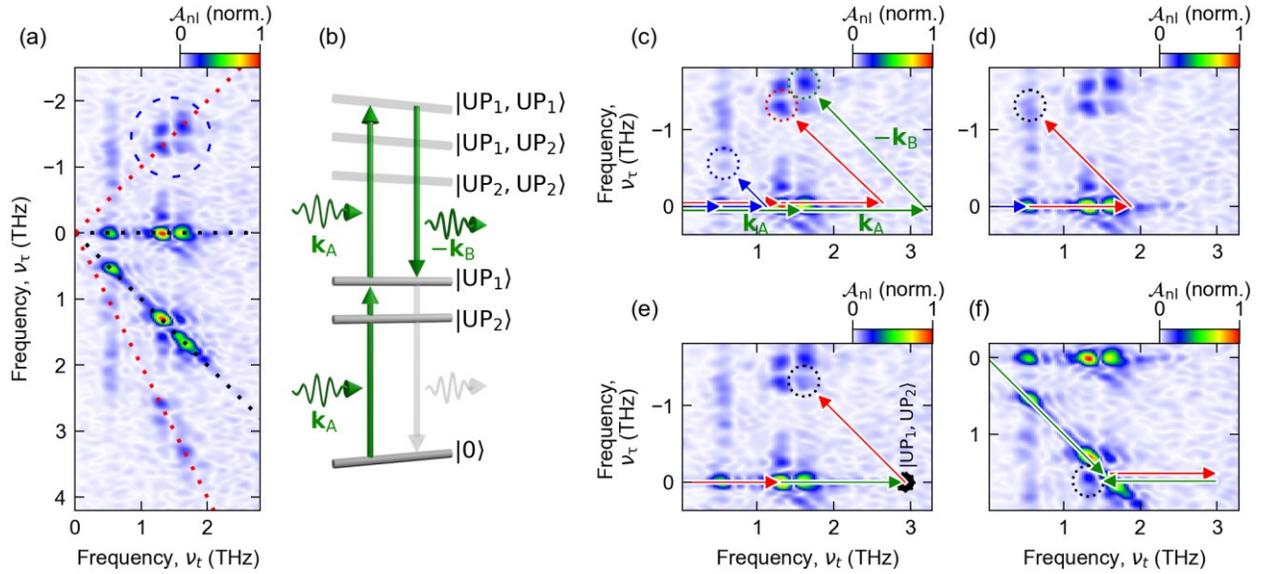

FIG. 3. (a) Experimental amplitude spectrum $\mathcal{A}_{nl}(\nu_t, \nu_\tau)$ of the time-domain data of Figure 2c, normalized to its peak value. Dashed lines as guides to the eye: Third-order processes with only one resonance. Dashed circle: Four-wave-mixing processes of $UP_1$ and $UP_2$. (b) Schematic of a four–wave mixing process. Two photons from field A with wave vectors $k_A$, resonant to $UP_1$ (left green waveforms), create a virtual excitation which is mixed down to the polariton frequency by emission of a photon into field B with wave vector $k_B$ (right green waveform). Reemission into the far field is illustrated by the light gray arrow. Other virtual levels $|UP_1, UP_2\rangle$, $|UP_2, UP_2\rangle$ involving the UP states are also shown, for completeness. (c-f) Liouville path analysis of nonlinear interactions: (c) Wave vector decomposition of the four-wave mixing processes of the $LP_1$ (blue arrows), $UP_2$ (red arrows) and $UP_1$ resonance (green arrows and wave vectors), individually. (d) Liouville path for a four-wave mixing process combining the nonlinear polarization of $LP_1$ and $UP_2$. (e,f) Four-wave mixing processes mixing the $UP_1$ and $UP_2$ resonances. The black dot in (e) marks the virtual level $|UP_1, UP_2\rangle$ as shown in panel b.

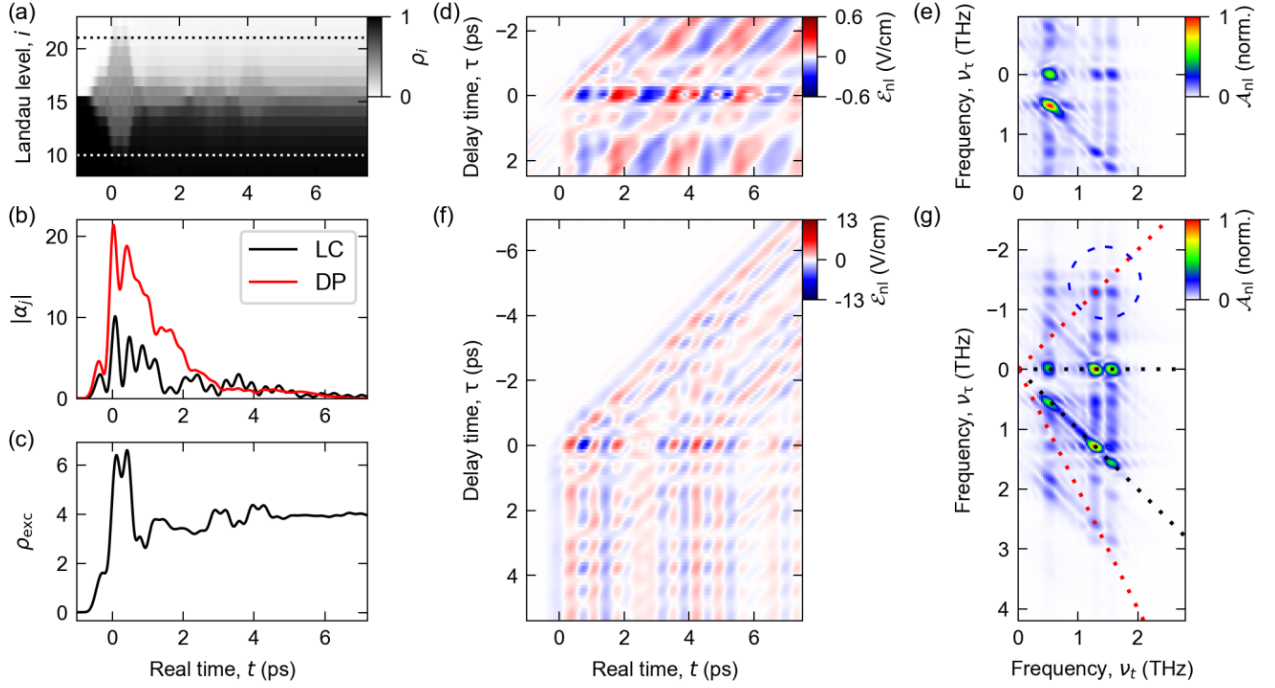

FIG. 4. Mean-field calculations. (a) Population of each Landau level for the excitation setting of Fig. 2c. Dashed lines: threshold energy for longitudinal optical phonon scattering (see text). (b) Population of the LC and DP cavity modes (see text) for $\tau = 0$ ps. (c) Electronic excitation $\rho_{\mathrm{exc}}$. (d) Calculated response $\mathcal{E}_{\mathrm{nl}}(t, \tau)$ considering only bandstructure effects. (e) Amplitude spectrum of the data of panel d, normalized to its peak value. (f) $\mathcal{E}_{\mathrm{nl}}(t, \tau)$ from our full theory including Coulomb effects. (g) Amplitude spectrum of the data of panel f, normalized to its peak value.

# Tailored subcycle nonlinearities of ultrastrong light-matter coupling

## Supplemental Material

J. Mornhinweg[1], M. Halbhuber[1], C. Ciuti[2], D. Bougeard[1], R. Huber[1,†], and C. Lange[1,3†]

[1]*Department of Physics, University of Regensburg, 93040 Regensburg, Germany*

[2]*Université de Paris, Laboratoire Matériaux et Phénomènes Quantiques,*

*CNRS, F-75013 Paris, France*

[3]*Fakultät Physik, Technische Universität Dortmund, 44227 Dortmund, Germany*

## Table of contents

# 1. Design of the light-matter coupled structure

The design of our ultrastrongly coupled structures is based on finite-element frequency-domain calculations in analogy to ref. [12]. The outer dimensions of the THz resonators are 37.5 μm by 30 μm, while the width of the metallic bars is 4 μm. They feature a double capacitive element with a gap of 2.5 μm and a width of 10 μm, in the center of which the electric near-field of the fundamental LC mode is enhanced by up to a factor of 8 with respect to the far-field (Fig. S1). The measured transmission spectrum of a bare resonator identifies the LC mode at a frequency of 0.81 THz, and the higher-order, dipolar (DP) mode at a frequency of 1.8 THz (Fig. S2a).

Our finite-element calculations can be extended by the gyrotropic response of the CR in the QWs, enabling the calculation of the transmission of the ultrastrongly coupled structure without free fit parameters [12]. Optimum agreement with the experimental spectra (Fig. S2b) is obtained for an electron doping density of $\rho = 2.1 \times 10^{12}$ cm$^{-2}$, per quantum well (Fig. S2c).

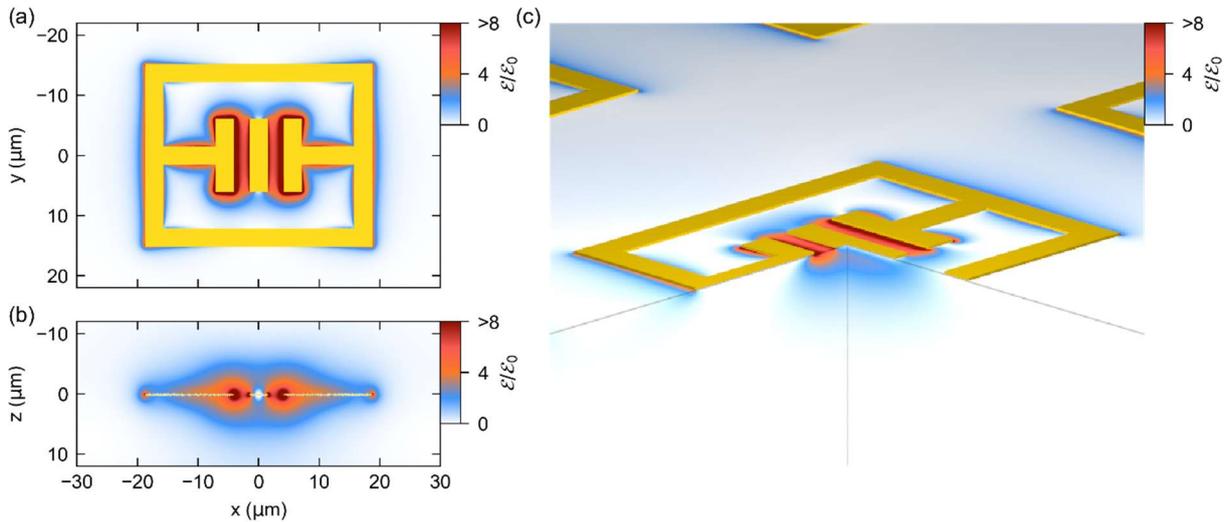

FIG. S1. Calculated enhancement of electric near field of the resonator structure. (a) Calculated electric near-field enhancement $\mathcal{E}/\mathcal{E}_0$, where $\mathcal{E}_0$ is the amplitude of the incident field, for the LC mode in the plane of the quantum well. (b) Near-field enhancement in the xz-plane along y = 0. (c) Three-dimensional cut-away view of $\mathcal{E}/\mathcal{E}_0$.

Moreover, the calculations can identify the individual contributions of the bare resonator modes to each of the coupled modes, in principle, by comparing the respective near-field distributions. Here, we restrict ourselves to a qualitative assessment. The near-field distribution of the $LP_1$ mode (Fig. S3a) matches the LC mode since it concentrates most of the field in the central gap region. The $UP_2$ mode, on the contrary, expels most of the field from the gap region (Fig. S3b), and is thus associated to the DP mode. Owing to its strong field enhancement in the gap, the $UP_1$ mode is linked to the LC mode (Fig. S3c).

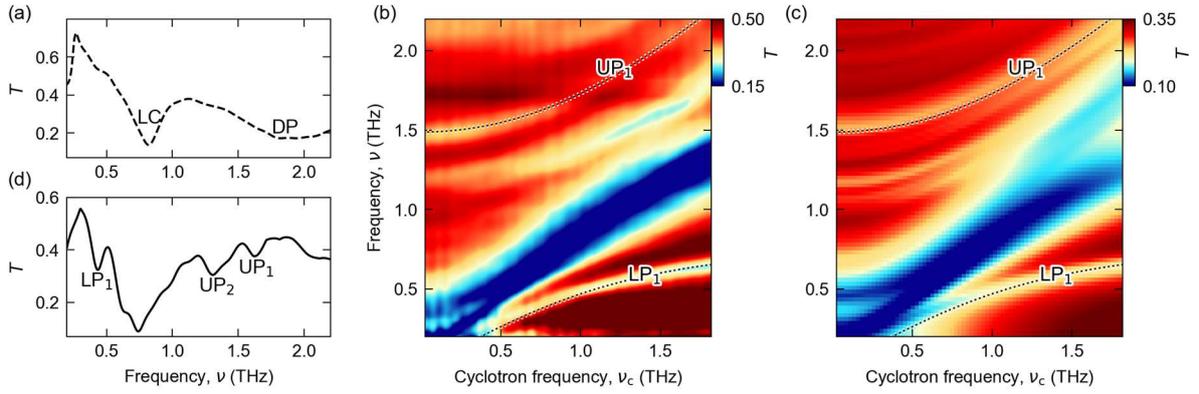

FIG. S2. Transmission spectra of the ultrastrongly coupled structure. (a) Measured spectrum of the bare resonator, for reference. (b) Measured spectra as a function of the CR frequency, $\nu_c$, as shown in the main text. Dashed lines: lower ($LP_1$) and upper ($UP_1$) polariton branches. Coupling strength: $\Omega_R^v/\omega_0 = 0.77$. (c) Calculated transmission spectra including the dashed lines of panel (b). (d) Transmission spectrum extracted from the data in (b), near the anti-crossing point at $\nu_c = 0.84$ THz.

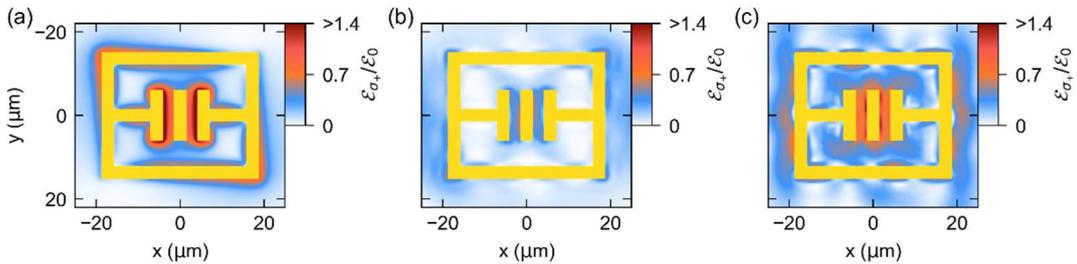

FIG. S3. Calculated right-circularly polarized near-field component $\mathcal{E}_{\sigma_+}/\mathcal{E}_0$, to which the cyclotron resonance selectively couples, whereby $\mathcal{E}_0$ is the far-field amplitude at $\nu_c = 0.84$ THz, (a) $\nu = 0.46$ THz for the $LP_1$, (b) $\nu = 1.30$ THz for the $UP_2$ and (c) $\nu = 1.73$ THz for the $UP_1$.

## 2. Experimental setup

We obtain the linear THz transmission using single-cycle THz waveforms (Fig. 1a, red waveform) generated in a ZnTe crystal ($\langle 110 \rangle$-cut, thickness of 1 mm) by optical rectification of femtosecond near-infrared pulses with a duration of 33 fs and a center wavelength of 807 nm, derived from a titanium-sapphire laser amplifier operating at a repetition rate of 3 kHz. The transmitted radiation is detected electro-optically in a second ZnTe crystal ($\langle 110 \rangle$-cut, thickness of 0.5 mm).

In order to record the nonlinear response we increase the peak amplitude of the first pulse, labelled A, to 1.3 kV/cm and add a second THz pulse, labelled B (Fig. 2a), delayed with respect to pulse A by a variable delay time $\tau$. Pulse B is generated using a tilted-pulse-front scheme and its amplitude is set to 2.5 kV/cm, corresponding to a Rabi frequency of $\Omega_R = 2\pi \times 2.7$ THz. Mechanical choppers, synchronized to sub-harmonics of the laser repetition rate, allow us to record all possible combinations of THz excitation with two pulses: both pulses incident on the sample, each pulse incident on the sample individually, as well as no THz illumination, for referencing.

## 3. Calibration of field amplitudes

We quantified the field strength of our THz pulses by measuring the power from our tilted-pulse-front THz source and calculating the corresponding peak amplitude, taking into account the THz waveform obtained by electro-optic sampling, as well as the beam profile measured with a pinhole, at the position of the sample. The resulting peak field amplitude of 1 MV/cm was attenuated in the experiment using filters and THz polarizers.

The parameters controlling the field amplitudes in our calculation were calibrated by comparing the density of electronic excitations in the theory and in the experiment. To this end, we calculated the absorption of the THz pulses, taking into account the linear transmission spectra of our structure within the spectral range of the three polariton resonances, and the corresponding spectral power density of the incident THz pulses. Factoring in the gap area of our THz resonators, where the strongest contribution to light-matter interaction prevails, we obtain an excitation density of $6.95 \times 10^{11}$ cm$^{-2}$ Landau electrons, per quantum well. Subsequently, the amplitude of the THz signal driving the cavity modes is scaled such that the density of excitations, $\rho_{\text{exc}} \times 2eB/\hbar$, matches the excitation density of the experiment. Note that here, the excitation parameter $\rho_{\text{exc}}$ is multiplied by the density of states of each Landau cylinder of $2eB/\hbar$ [S1].

Finally, we allow for different coupling of the two cavity modes to the incident far field by an individual coupling factor $\kappa_j$, whereby $\kappa_{\text{LC}} = 1$ and $\kappa_{\text{DP}} = 1.5$ for the LC and DP modes.

## 4. Calculation of the coherent Rabi frequency

While the vacuum Rabi frequency of our structure, $\Omega_R^v$, can be determined by the frequencies of the polariton branches measured in equilibrium, the Rabi frequency of the coherent drive, $\Omega_R$, depends on the external field amplitude. We determine $\Omega_R$ by calculating the field distribution of the coupled system in the plane of the QW by finite element frequency-domain calculations. Subsequently, the local near-field waveforms induced by the far-field THz transient can be calculated by multiplying the complex-valued frequency-domain data with the corresponding spectrum of our THz transient, and back-transforming the result into the time domain. Since the phases of the near-field components do not vary strongly within the QW plane, the resulting near-field waveforms are all of very similar shape up to an amplitude scaling factor. Averaging these waveforms within the capacitive gap region of our THz resonators, we obtain a peak near-field amplitude of 1.7 kV/cm for excitation with a peak far-field amplitude of 2.5 kV/cm, and correspondingly, a coherent Rabi frequency of $\Omega_R = 2\pi \times 2.7$ THz.

## 5. Two-dimensional nonlinear amplitude spectrum for $\nu_c = 0$ THz

Without external magnetic field, the two-dimensional nonlinear amplitude spectrum corresponding to the time-domain data of Fig. 2b displays distinct resonances. In comparison to the data at the anti-crossing point, here, the LP resonance is absent and correspondingly, all nonlinear interactions result from the $UP_1$ and $UP_2$ resonances. Moreover, while the amplitudes of the pump-probe signals of these two resonances are comparable, interactions between $UP_1$ and $UP_2$ are strongly reduced owing to the absence of ultrastrong coupling (Fig. S4, blue circle).

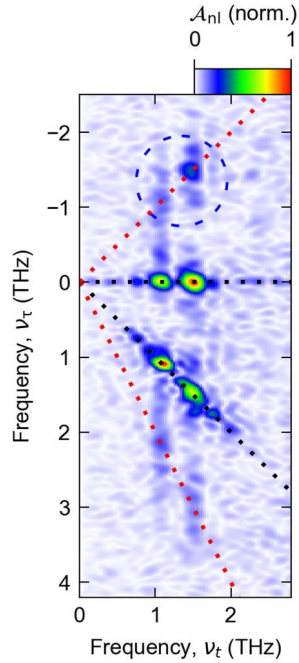

FIG. S4. Measured amplitude spectrum $\mathcal{A}_{nl}(\nu_t, \nu_\tau)$ of the time-domain data of Figure 2b, normalized to its peak value. Dashed lines as a guide to eye: Third-order processes involving only a single resonance. Dashed circle: Four-wave-mixing processes of the $UP_1$ and $UP_2$.

# 6. Quantitative comparison and experimental noise

For a direct comparison of experimental data and simulations, we extract cuts of the amplitude spectra $\mathcal{A}_{nl}(\nu_t, \nu_\tau)$ for different paths within the 2D spectrum (Fig. S5). For $\nu_\tau = 0$ (Fig. S5b), the frequencies and amplitudes of the pump-probe responses of the three polariton resonances match almost perfectly between experiment and theory. For the cut along $\nu_\tau = \nu_t$ (Fig. S5c), the resonances match as well. Here, some deviation is caused by spectral broadening of the $UP_2$. The cuts along $\nu_\tau = 1.3$ THz and $\nu_\tau = 1.6$ THz (Fig. S5d,e) again show a good agreement of measured and calculated resonance frequencies. For completeness we also show cuts along $\nu_t = 0.55$ THz, for 1.3 THz and 1.6 THz (Fig. S5f-h).

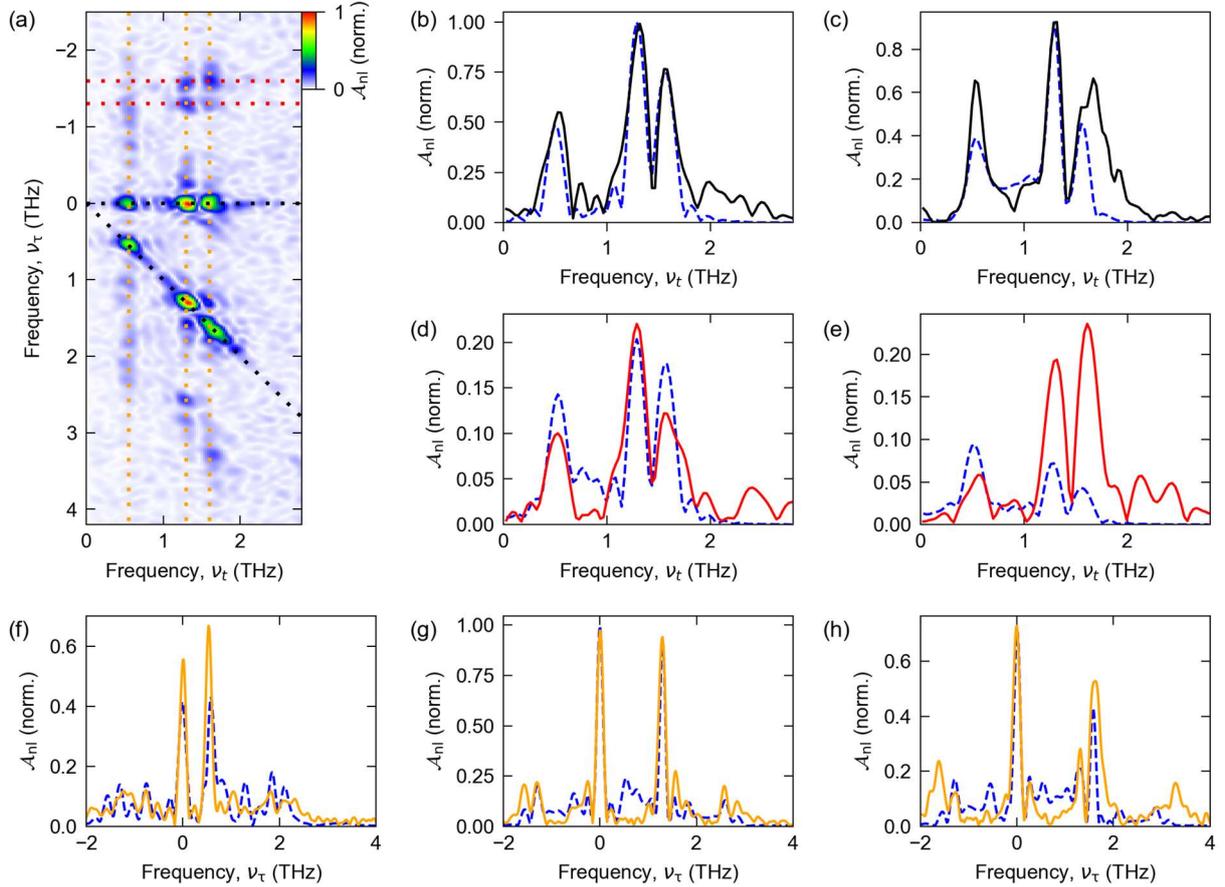

FIG. S5. (a) Amplitude spectrum $\mathcal{A}_{nl}(\nu_t, \nu_\tau)$ of the time-domain data of Figure 2c, normalized to its peak value. (b-h) Amplitude spectra comparing the experimental data (solid curves) and theoretical calculations (dashed blue curves, see also Figure 4), extracted along the dashed lines shown in panel (a) of the same color. (b) Spectrum extracted for $\nu_\tau = 0$, (c) for $\nu_\tau = \nu_t$, (d) for $\nu_\tau = 1.3$ THz, (e) for $\nu_\tau = 1.6$ THz, (f) for $\nu_t = 0.55$ THz, (g) for $\nu_t = 1.3$ THz and (h) for $\nu_t = 1.6$ THz.

Moreover, we assess the experimental noise level by calculating the standard error of the nonlinear spectra. For each time tuple ($t$, $\tau$) of the experimental time-domain data, $n = 20$ individual data points, $\mathcal{E}_{\text{nl},i}$, $i = 1..n$, were measured, resulting in $n$ independent nonlinear spectra $\mathcal{A}_{\text{nl},i}$, $i = 1..n$. Averaging these spectra results in the spectra discussed in the manuscript, $\mathcal{A}_{\text{nl}} = \langle \mathcal{A}_{\text{nl},i} \rangle_i$ (Fig. S6a), where $\langle ... \rangle_i$ denotes the average of all samples $i = 1..n$. On the other hand, calculating their standard deviations for each frequency tuple ($\nu_t$, $\nu_\tau$) yields information on the spectral noise (Fig. S6b,c). From these data, a normalized, frequency-averaged standard error of $\langle \sigma_n \rangle_{\nu_t,\nu_\tau} / \max(\mathcal{A}_{\text{nl}}) = 1.2\%$ was determined. Here, $\sigma_n = \sigma(\mathcal{A}_{\text{nl},i})/\sqrt{n}$ is the standard error calculated for each frequency tuple, and $\langle ... \rangle_{\nu_t,\nu_\tau}$ denotes the averaging along the entire range of frequencies $\nu_t$ and $\nu_\tau$. Moreover, the standard error for $\mathcal{E}_{\text{nl},i}(t, \tau)$ allows for a quantitative evaluation of the beating pattern of the UP$_1$ and UP$_2$ resonances (compare to Fig. 2b,c) and the additional modulation caused by the contribution by the LP$_1$ resonance (Fig. S7).

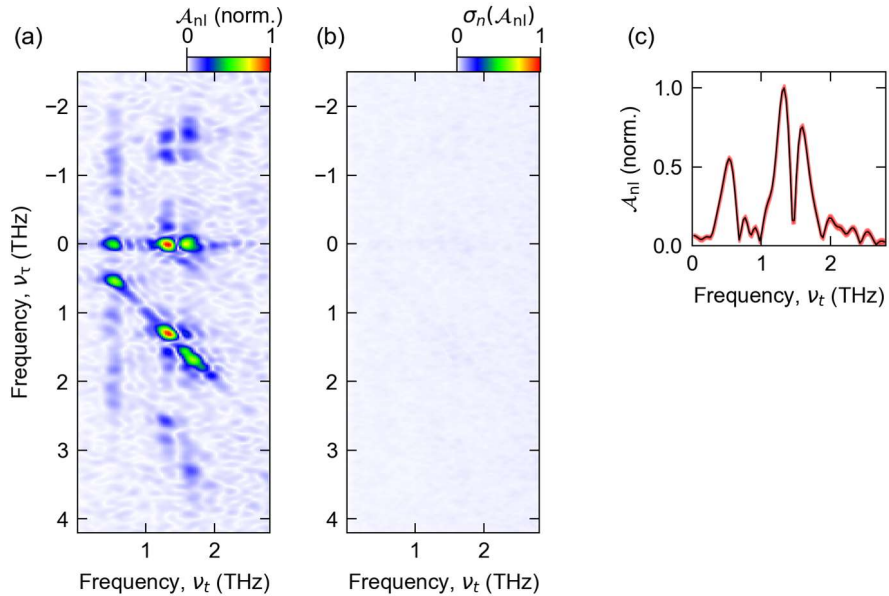

FIG. S6. (a) Experimental nonlinear amplitude spectra $\mathcal{A}_{\text{nl}}$ of Fig. 3(a), normalized to their peak value, and (b) standard error $\sigma_n(\mathcal{A}_{\text{nl}})$, shown on the same scale as $\mathcal{A}_{\text{nl}}$ in panel a. (c) $\mathcal{A}_{\text{nl}}$ at $\nu_\tau = 0$ (black curve), and error band, $\mathcal{A}_{\text{nl}} \pm \sigma_n(\mathcal{A}_{\text{nl}})$ (red shade).

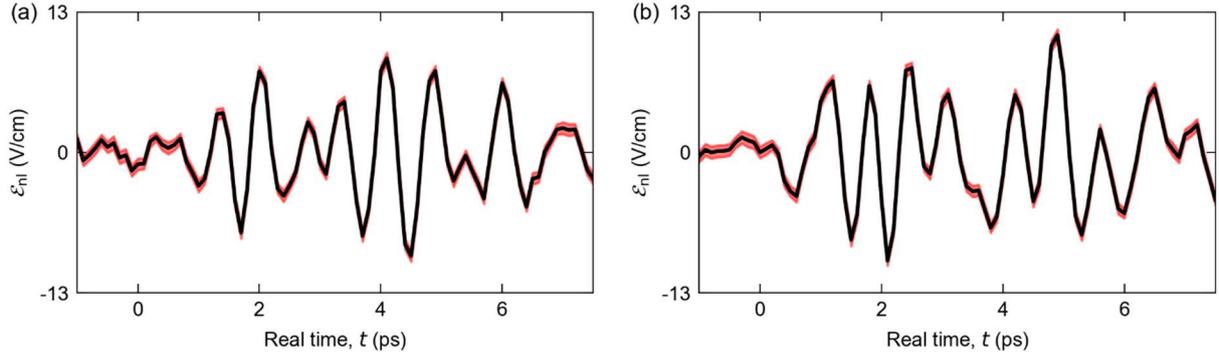

FIG. S7. Measured nonlinear electric field $\mathcal{E}_{nl}$ of Fig. 2, averaged from $\tau$ = -0.1 ps to $\tau$ = 0.1 ps. (a) $\mathcal{E}_{nl}$ for $\nu_c$ = 0 THz (Fig. 2b), showing the beating pattern of the $UP_1$ and $UP_2$ resonances. (b) $\mathcal{E}_{nl}$ for $\nu_c$ = 0.84 THz (Fig. 2c), showing an additional modulation caused by the contribution by the $LP_1$ resonance. Both plots show the standard error as a red band spanning the range $\mathcal{E}_{nl} \pm \sigma_n(\mathcal{E}_{nl})$.

# 7. Excitation with a peak field amplitude of 5.6 kV/cm

The nonlinearities discussed in the main text exhibit a high degree of coherence owing to the narrowband polariton resonances of our structures. We test the limits of this coherence by performing an additional measurement with an approximately doubled field amplitude of pulse B of 5.6 kV/cm, while the characteristics of pulse A are not altered. In this setting, the time-domain data exhibit a strong reduction of the amplitude of the oscillations of $\mathcal{E}_{nl}$ along the $\tau$ axis as compared to the data of Fig. 2, and the dynamics is governed by a slowly decaying background resulting from incoherent population (Fig. S8a). Correspondingly, the spectral amplitude $\mathcal{A}_{nl}$ is structured predominantly by pump-probe signatures at frequencies $(\nu_j, 0)$ and $(\nu_j, \nu_j)$, where $j \in \{LP_1, UP_1, UP_2\}$ is the polariton index, while four-wave mixing signals at $(\nu_j, -\nu_j)$ and $(\nu_j, 2\nu_j)$ contribute only weakly (Fig. S8b). The significantly larger amplitude of pulse B as compared to pulse A leads to a preference of third-order processes including two photons from pulse B, such as the pump-probe signatures at frequencies of $(\nu_j, 0)$.

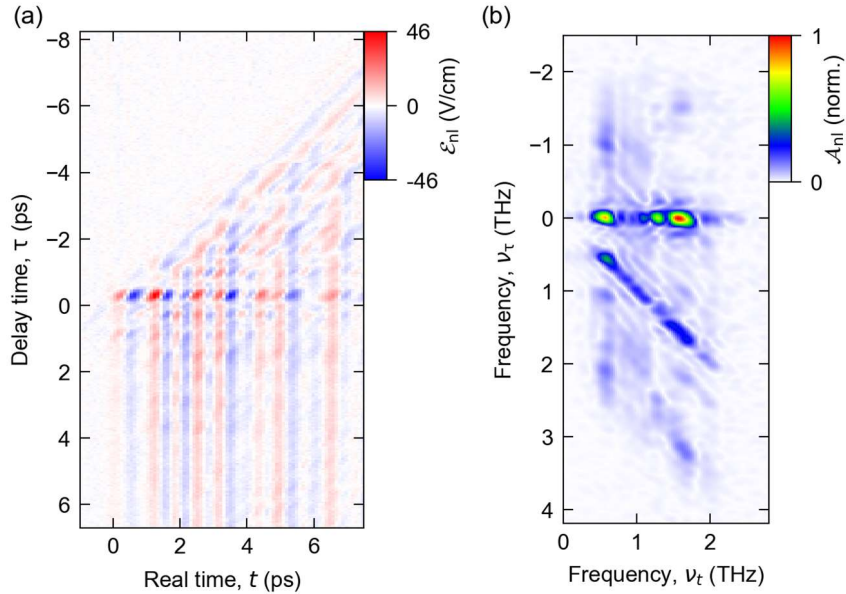

FIG. S8. (a) Measured nonlinear polarization response $\mathcal{E}_{nl}(t, \tau)$ for a peak field amplitude of pulse B of 5.6 kV/cm. (b) Spectral amplitudes $\mathcal{A}_{nl}(\nu_t, \nu_\tau)$ obtained from the time-domain data of panel (a), normalized to its peak value.

The calculations for the data of Fig. 2 have demonstrated that the distribution of excited electrons along the Landau fan covers an energy range which already slightly extends beyond the threshold above which scattering of Landau electrons with longitudinal optical phonons of GaAs becomes possible [29]. We account for phonon scattering of these electrons by a short dephasing time of $T_2 = 100$ fs applied for Landau electrons excited to Landau levels outside of the range $E_F \pm \hbar\omega_{LO}/2$, marked by the dashed white lines in Fig. 4a. Here, $E_F$ is the Fermi level in equilibrium, and $\hbar\omega_{LO}$ is the energy of the LO phonon. This choice is motivated by the high, albeit not perfect symmetry of the distribution of excited Landau electrons, leading to the simultaneous availability of empty Landau states at $E_F - \hbar\omega_{LO}/2$ and occupied ones at $E_F + \hbar\omega_{LO}/2$ for sufficiently strong excitation, and, correspondingly, opening up the phonon scattering channel. The calculation, shown in Fig. S9a, accurately reproduces the predominantly incoherent pump-probe nonlinearities observed in the experimental data of Fig. S8, and shows that a significant fraction of the total population is excited beyond the threshold for scattering (Fig. S9b).

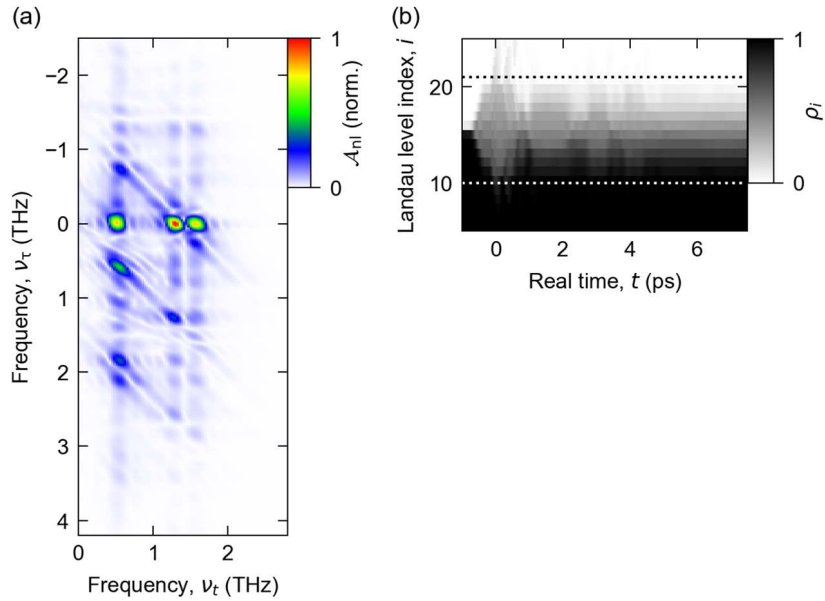

FIG. S9. (a) Calculated nonlinear spectrum $\mathcal{A}_{nl}(t, \tau)$ for a peak field amplitude of pulse B of 5.6 kV/cm, normalized to its peak value. (b) Time-dependent population density $\rho_{el}$ for $\tau = 0$ ps. The two dashed lines mark the energy threshold for scattering with the longitudinal optical phonon of GaAs.

## 8. Nonlinear response of a structure based on a single QW

For reference, we measured the response of a sample with a single quantum well (thickness of 10 nm, n-doped with an electron density of $6.3 \times 10^{11}$ cm$^{-2}$) and a coupling strength of $\Omega_R^v/\omega_c = 0.15$. The transmission as a function of the cyclotron frequency is shown in Fig. S10, where two distinct polariton branches are visible. At the anti-crossing point, the frequencies of the lower and upper polariton modes are 0.7 THz and 0.95 THz, respectively.

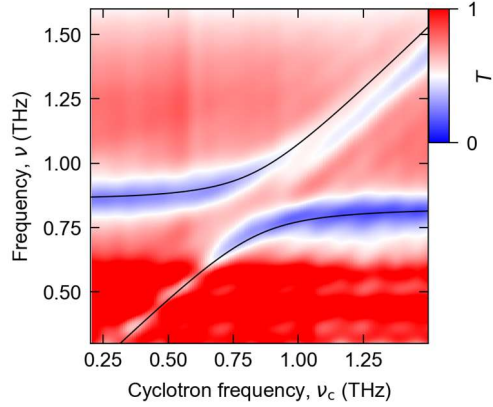

FIG. S10. Measured transmission spectra of a single-QW structure as a function of the cyclotron resonance frequency, $\nu_c$. The lines mark the lower (LP$_1$) and upper (UP$_1$) polariton branches with a splitting corresponding to a coupling strength of $\Omega_R^v/\omega_c = 0.15$.

Performing the nonlinear measurement at the anti-crossing point at $\nu_c = 0.8$ THz, we observe long-lived oscillatory features, in the time domain, modulated by several frequency components along the $\tau$ axis (Fig. S11a). Two-dimensional Fourier transformation extracts the nonlinear amplitude spectrum (Fig. S11b), which reveals pump-probe (PP) signatures for $\nu_\tau = 0$ THz, at the frequencies of the two polariton resonances. Near $\nu_\tau = -0.8$ THz and $-1.6$ THz, groups of four (4WM) and six-wave mixing (6WM) signatures are located (see Fig. S11c,d for a magnified view). Near $\nu_\tau = 2.2$ THz, 3.2 THz, and 4.6 THz, four, six and even a faint eight-wave mixing signature (8WM) emerge.

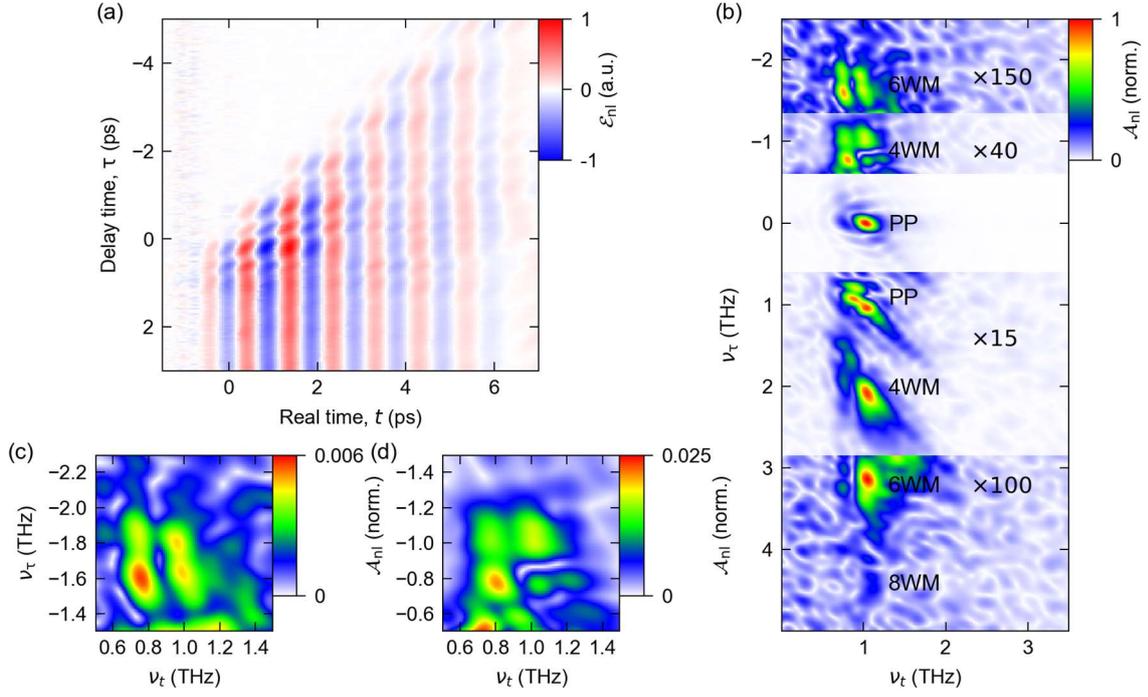

FIG. S11. Experimental nonlinear polarization response of a single-QW sample for a peak field amplitude of 9 kV/cm, normalized to its peak value. (a) Time-domain data $\mathcal{E}_{nl}(t, \tau)$. (b) Spectral amplitudes $\mathcal{A}_{nl}(\nu_t, \nu_\tau)$ obtained from the time-domain data of panel a. For each vertical segment, the data are scaled by the specified factor, for better visibility. (c) Close-up of a group of six-wave and (d) four-wave mixing signatures.

# 9. Combinatorial model of the nonlinear response of multiple polaritons

In the main text, we have presented a Liouville path analysis of a selection of the strongest nonlinear signals and their wave vector dissection. The full spectrum of all possible third-order nonlinearities of the three polariton branches results from considering all combinations of the corresponding three wave vectors, taking into account both fields, A and B, and both processes: absorption and emission. Figure S12 shows all pump-probe (black circles) and all four-wave-mixing processes (red circles). Filled circles indicate processes resulting in a nonlinear polarization oscillating at the frequency of one of the polariton resonances, leading to strong emission into the far field. Open circles, on the contrary, belong to off-resonant polarization components and do not contribute to the total emission significantly. Moreover, since some locations in frequency space may be reached by more than one Liouville path, we visualize the degeneracy by the size of each circle.

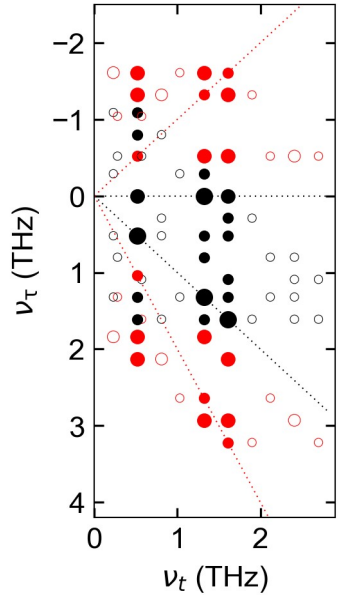

FIG. S12. All combinatorially possible nonlinear resonances of $\mathcal{A}_{nl}(\nu_t, \nu_\tau)$ for third-order nonlinear interactions between the LP, the $UP_1$ and the $UP_2$ resonances, colored black and red for pump-probe and four-wave-mixing processes, respectively. The size of the dots scales with the number of Liouville paths leading to the center position of the dot. See text for more details.

# 10. Subcycle quantum model of ultrastrong light-matter coupling

**Bosonic multi-mode model of ultrastrong coupling**

We reproduce the experimental data by a specifically developed mean-field theory which treats light-matter interaction in the ultrastrong coupling regime beyond the rotating-wave approximation. For the resonator, two effective modes at center frequencies of $\nu_{LC} = 0.86$ THz and $\nu_{DP} = 1.5$ THz are considered, while the electronic system is implemented by a density matrix formalism which accounts for the coherent polarization and incoherent population of a Landau fan of $10^2$ individual levels.

The corresponding Hamiltonian is written as

$$\hat{H} = \hat{H}_{\text{cavity}} + \hat{H}_{\text{Landau}} + \hat{H}_{\text{int}} + \hat{H}_{\text{dia}} + \hat{H}_{\text{ext}}, \tag{1}$$

and contains several contributions as described in the following. The cavity modes are contained in

$$\hat{H}_{\text{cavity}} = \sum_j \hbar \omega_j \hat{a}_j^\dagger \hat{a}_j, \tag{2}$$

where $\omega_j$ is the frequency, and $\hat{a}_j$ is the annihilation operator of the $j$-th mode ($j \in \{\text{LC}, \text{DP}\}$).

In a first step, excitations of the Landau system are implemented in the bosonic limit, and

$$\hat{H}_{\text{Landau}} = \hbar \omega_c \hat{b}^\dagger \hat{b}. \tag{3}$$

Here, $\omega_c$ denotes the cyclotron frequency and $\hat{b}$ is the annihilation operator of the Landau excitations. Light-matter coupling including anti-resonant interaction terms is introduced by

$$\hat{H}_{\text{int}} = \hbar \sum_j \Omega_{R,j}^V (\hat{a}_j + \hat{a}_j^\dagger)(\hat{b} + \hat{b}^\dagger), \tag{4}$$

where $\Omega_{R,j}^V$ is the vacuum Rabi frequency of $j$-th cavity mode coupling to the CR. The blue-shift of the cavity modes by diamagnetic interactions is accounted for by

$$\hat{H}_{\text{dia}} = \hbar \sum_j D_j (\hat{a}_j + \hat{a}_j^\dagger)^2, \tag{5}$$

where $D_j = \frac{(\Omega_{R,j}^V)^2}{\omega_c}$.

Coupling of the cavity modes to the THz far-field is contained in $\hat{H}_{\text{ext}}$.

We use Heisenberg's equation of motion for each of these operators analogously to

$$\frac{d\hat{A}(t)}{dt} = -\frac{i}{\hbar}[\hat{A}(t),\hat{H}], \tag{6}$$

where $\hat{A}(t)$ is a bosonic operator. Applying corresponding bosonic commutation relations, we derive the time-dependent differential equations:

$$\frac{d\hat{a}_j(t)}{dt} = -i\omega_j\hat{a}_j(t) - i\Omega^v_{R,j}[\hat{b}(t)+\hat{b}^\dagger(t)] - i2D_j[\hat{a}_j(t)+\hat{a}_j^\dagger(t)] - \frac{i}{\hbar}[\hat{a}_j(t),\hat{H}_{\text{ext}}], \tag{7}$$

$$\frac{d\hat{b}(t)}{dt} = -i\omega_c\hat{b}(t) - i\sum_j \Omega^v_{R,j}[\hat{a}_j(t)+\hat{a}_j^\dagger(t)]. \tag{8}$$

We use a mean-field approach $\alpha_j \equiv \langle\hat{a}_j\rangle(t)$ and $\beta \equiv \langle\hat{b}\rangle(t)$, omit the explicit temporal dependence for simplicity, and introduce phenomenological dephasing rates $\gamma_j$ for each mode:

$$\frac{d}{dt}\alpha_j = -i\omega_j\alpha_j - \gamma_j\alpha_j - i\Omega^v_{R,j}[\beta+\beta^*] - i2D_j[\alpha_j+\alpha_j^*] + \kappa_j\sqrt{\gamma_j}\,\mathcal{E}_{\text{THz}}(t), \tag{9}$$

$$\frac{d}{dt}\beta = -i\omega_c\beta - i\sum_j \Omega^v_{R,j}[\alpha_j+\alpha_j^*]. \tag{10}$$

The external THz field $\mathcal{E}_{\text{THz}}(t)$ is coupled to each cavity mode by an individual coupling constant $\kappa_j$ accounting for its dipole moment, and a factor of $\sqrt{\gamma_j}$ which results from the fluctuation-dissipation theorem.

**Density matrix formalism of the multi-level Landau fan**

In a second step, we introduce the full complexity of the multi-level Landau fan including its non-equidistant energy progression by a density matrix approach for $N = 10^2$ levels:

$$\hat{\mathcal{H}}_\rho = \begin{pmatrix} \hbar\omega_1 & \mu_{12}e^{-i\omega_{12}t} & \cdots & 0 \\ \mu_{12}e^{i\omega_{12}t} & \ddots & \ddots & \vdots \\ \vdots & \ddots & \ddots & \vdots \\ 0 & \cdots & \cdots & \hbar\omega_N \end{pmatrix}, \tag{11}$$

$$\rho = \begin{pmatrix} \rho_{11} & \cdots & \rho_{1N} \\ \vdots & \ddots & \vdots \\ \rho_{N1} & \cdots & \rho_{NN} \end{pmatrix}. \tag{12}$$

The diagonal of $\hat{\mathcal{H}}_\rho$ contains the eigenenergies of the $j$-th Landau level, $\hbar\omega_j$, while the adjacent minor diagonal is populated by coupling terms of neighboring Landau levels according to the selection rules as well as the driving field, leading to

$$\mu_{mn} = d_{mn} \times \sum_{j}(\alpha_j + \alpha_j^*)\Omega_{R,j}^v. \tag{13}$$

Here, the sum of the real parts of the cavity fields, scaled by their respective vacuum Rabi frequency, is multiplied by the dipole moment, acting as the driving term for the transition. The dipole moments scale according to $d_{mn} = el_0\sqrt{m}$ for the transition from Landau level $m$ to $n$, whereby $e$ is the elementary charge, $l_0 = \sqrt{\hbar/eB}$ is the magnetic length, and $B$ is the static magnetic bias field.

The resulting density matrix $\rho$ contains the population density of Landau level $m$ on its diagonal, $\rho_{mm}$, while the off-diagonal entries $\rho_{mn} = \rho_{nm}^*$ describe the coherent polarization between Landau levels $m$ and $n$. The temporal evolution of the density matrix is calculated by solving the von Neumann equation,

$$-i\hbar \frac{\partial \rho}{\partial t} = [\hat{\mathcal{H}}, \rho] - i\gamma \circ \rho, \tag{14}$$

whereby $\gamma$ is a damping matrix which phenomenologically implements dephasing for each element of $\rho$ individually, by the element-wise product, " $\circ$ ".

We apply this formalism to the theory of ultrastrong coupling by replacing the polarization term of Eq. (9), $\beta(t) + \beta^*(t)$, by the corresponding polarization field of the multi-level Landau system,

$$P_L(t) = \sum_{m,n} \rho_{mn}(t) d_{mn}. \tag{15}$$

Omitting again the explicit time dependence, the modified version of Eq. (9) finally assumes the form

$$\frac{d}{dt}\alpha_j = -i\omega_j\alpha_j - \gamma_j\alpha_j - i\Omega_{R,j}^v P_L - i2D_j[\alpha_j + \alpha_j^*] + \sqrt{\gamma_j}\kappa_j\,\mathcal{E}_{THz}(t). \tag{16}$$

Solving the system of differential equations, we obtain the time-dependent density matrix, $\rho$, and the cavity fields, $\alpha_j(t)$. The total field corresponding to the field measured in the experiment is obtained by superimposing the fields reradiated by both cavity modes, and the incident THz transient, $\mathcal{E}_{THz}$:

$$\mathcal{E}_{measured}(t) = \mathcal{E}_{THz}(t) - \sum_{j}\sqrt{\gamma_{cav,j}}\,\alpha_j \tag{17}$$

The nonlinear response $\varepsilon_{nl}$ is calculated analogously to the measurement by simulating all three possible combinations of both THz pulses A and B, only pulse A, and only pulse B exciting the structure.

**Implementation of nonlinearities**

Our model considers three sources of nonlinearities, including the shape of the conduction band of the GaAs quantum wells as well as dynamical Coulomb correlations influencing both the Landau level energies as well as the transition dipole matrix elements.

The non-parabolic shape of the conduction band leads to a non-equidistant progression of the Landau level frequencies which can be modelled as

$$\omega_j = \sqrt{\frac{\omega_{np}^2}{4} + \frac{\omega_{np} eB}{m^*(j + 0.5)}}, \qquad (18)$$

where $j$ is the Landau level index, the effective mass is given by $m^* = 0.066\, m_e$, $e$ is the elementary charge, $B$ is the static magnetic bias field, and $\omega_{np} = 2\pi \times 237$ THz is a fit parameter for GaAs [31]. For the doping density of our sample of $1.75 \times 10^{12}$ cm$^{-2}$ and a magnetic field of 2.3 T, at the anti-crossing point, the filling factor is $n_{LL} = 15.73$. In this configuration, the transition frequency in the vicinity of the Landau level is $\omega_c = 0.868$ THz, while the next-higher and next-lower transitions relevant for strong-field excitation are centered at 0.874 THz and 0.862 THz, respectively. Owing to this small deviation of < 1%, the calculation implementing only effects related to this non-parabolic band shape yields small nonlinearities of an amplitude of 0.07 % of the incident field (see main text).

On the other hand, strong, non-perturbative excitation of the a massively many-body Landau system has been shown to lead to dynamical Coulomb correlations and corresponding sizeable nonlinearities [29]. We implement these effects by dynamically adjusting the Landau level eigenenergies and dipole moments during excitation, depending on the instantaneous degree of excitation of the electronic system. As a measure of this excitation, we introduce

$$\rho_{exc}(t) = \sum_m \frac{1}{2} \left| \text{diag}(\rho(t)) - \text{diag}(\rho_0) \right|_m l_m, \qquad (19)$$

$$l_m = \begin{cases} m - j_f & \text{for} \quad m < j_f \\ m - j_f + 1 & \text{for} \quad m > j_f \end{cases}, \qquad (20)$$

considering both the number of excited Landau electrons as well as their distance in Landau level space relative to the equilibrium distribution, $\rho_0$, and equilibrium Landau level index at the Fermi level, $j_f$. This parameter can be intuitively interpreted for the hypothetical case of a perfectly parabolic conduction band, where an infinitely extended Landau fan behaves as a perfect harmonic oscillator and $\rho_{\text{exc}}$ is equal to the number of electronic excitations of the bosonic version of our theory, $|\beta|^2$.

With this approach, we renormalize the Landau level frequencies of Eq. (18), and correspondingly, the transitions energies, by a dynamical scaling factor $U_e$,

$$\omega_j \rightarrow \omega_j \times (1 + U_e \rho_{\text{exc}})^{-1}. \tag{21}$$

Analogously, we rescale the values of all dipole moments of Eq. (13) by a scaling factor $U_d$,

$$d_{mn} \rightarrow d_{mn} \times (1 + U_d \rho_{\text{exc}})^{-1}. \tag{22}$$

As discussed in the main text, only the combination of all three nonlinearities allows for a quantitative description of the experimental data ($U_e = 0.016$, $U_d = 0.064$). The individual influence of each of these nonlinearities is discussed in the following chapter.

**Diagonalization of the light-matter coupling Hamiltonian**

To obtain the fits of the upper and lower polariton as shown in Figure 1b, we use the bosonic light-matter coupling Hamiltonian:

$$\widehat{H} = \sum_j \hbar \omega_j \hat{a}_j^\dagger \hat{a}_j, + \hbar \omega_c \hat{b}^\dagger \hat{b} + \hbar \sum_j \Omega_{R,j}^v (\hat{a}_j + \hat{a}_j^\dagger)(\hat{b} + \hat{b}^\dagger) + \hbar \sum_j D_j (\hat{a}_j + \hat{a}_j^\dagger)^2. \tag{23}$$

The light-matter Hamilton operator can be expressed in terms of normal-mode polariton operators, as we only fit the coupling to the LC mode, $\{\hat{p}_i, \hat{p}_i^\dagger\}_{i=\text{LP,UP}}$, whereby $\hat{p}_i = w_i \hat{a}_{\text{LC}} + x_i \hat{b} + y_i \hat{a}_{\text{LC}}^\dagger + z_i \hat{b}^\dagger$, via the Bogoliubov transformation. The coefficients $(w_i, x_i, y_i, z_i)$ are obtained by diagonalising the corresponding Hopfield matrix [S2],

$$M = \begin{pmatrix} \omega_{\text{LC}} + 2D & \Omega_R^v & 2D & \Omega_R^v \\ \Omega_R^v & \omega_c & \Omega_R & 0 \\ -2D & -\Omega_R^v & -\omega_{\text{LC}} - 2D & -\Omega_R^v \\ -\Omega_R^v & 0 & -\Omega_R^v & -\omega_c \end{pmatrix}.$$

## 11. Switch-off analysis of nonlinear contributions

In this section, we analyze the influence of each of the three sources of nonlinearities on the results obtained for the full calculation. With only the non-parabolic bandstructure considered, the nonlinear field $\mathcal{E}_{nl}$ exhibits coherent modulations in qualitative similarity to the experiment but a far lower amplitude of only 0.6 V/cm as compared to the measured amplitude of 13 V/cm (Fig. S13a). Moreover, the modulation of the calculated data is clearly slower and, correspondingly, the spectral data exhibit most of the spectral weight on the LP resonance (Fig. S13b). Adding only energy rescaling to the extent that is also used for the full calculation ($U_e = 0.016$) leads to a nonlinear amplitude matching that of the experiment, yet still emphasizing low-frequency components (Fig. S13c,d). Similar results are obtained when including non-parabolic effects combined with only the rescaling of the transition dipole moments ($U_d = 0.064$, Fig. S13e,f). Switching on both Coulomb-related nonlinearities but neglecting the non-parabolic shape of the conduction band, the spectral weight is shifted to the upper polariton resonances, and a very good match with the experiment is obtained (Fig. S13g,h). The best agreement is found for a combination of all three nonlinearities, as discussed in the main text and Fig. 4.

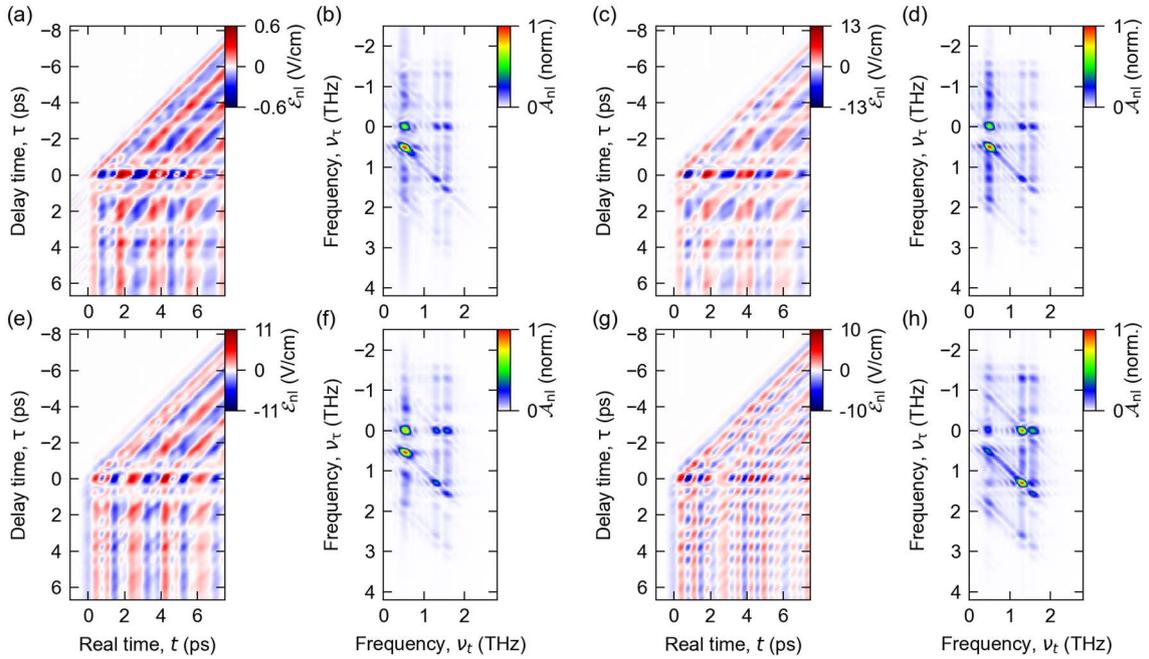

FIG. S13. Switch-off analysis of different nonlinearities. (a,b) Time-domain calculation and amplitude spectra, normalized to its peak value, for $U_d = U_e = 0$. (c,d) Calculation for $U_d = 0$. (e,f) Calculation for $U_e = 0$. (g,f) Calculation for equally spaced Landau levels.